\documentclass[runningheads]{llncs}

\usepackage[title]{appendix}
\usepackage{graphicx}
\usepackage{comment}
\usepackage{graphicx}
\usepackage{subfig}
\usepackage{lipsum}
\usepackage{hyperref}
\usepackage{comment}
\usepackage{tabularx}
\usepackage{booktabs}
\usepackage{caption}
\usepackage{graphicx}
\newcommand{\squeezeup}{\vspace{-2.5mm}}
\begin{document}
\title{ Assessing Sentiment of the Expressed Stance on Social Media}
% how the stance is being expressed in the social media 
%\titlerunning{Abbreviated paper title}
% If the paper title is too long for the running head, you can set
% an abbreviated paper title here
%
\author{Abeer Aldayel\and
Walid Magdy}
\authorrunning{A. Aldayel and W.Magdy.}
% First names are abbreviated in the running head.
% If there are more than two authors, 'et al.' is used.
%
\institute{School of Informatics. The University of Edinburgh, Edinburgh, UK \\
\email{a.aldayel@ed.ac.uk and wmagdy@inf.ed.ac.uk}\\
}
\maketitle              % typeset the header of the contribution
\begin{abstract}
Stance detection is the task of inferring viewpoint towards a given topic or entity either being supportive or opposing. One may express a viewpoint towards a topic by using positive or negative language. This paper examines how the stance is being expressed in social media according to the sentiment polarity. There has been a noticeable misconception of the similarity between the stance and sentiment when it comes to viewpoint discovery, where negative sentiment is assumed to mean against stance, and positive sentiment means in-favour stance. To analyze the relation between stance and sentiment, we construct a new dataset with four topics and examine how people express their viewpoint with regards these topics. We validate our results by carrying a further analysis of the popular stance benchmark SemEval stance dataset. Our analyses reveal that sentiment and stance are not highly aligned, and hence the simple sentiment polarity cannot be used solely to denote a stance toward a given topic.
%Overall the stance tends to have a negative sentiment to express the opposing or supporting viewpoint to an entity.  We discuss the practical implication of the situation where sentiment has been used instead of stance to evaluate the public opine toward a topic or entity. 
\\\textbf{This is a preprint of an article accepted for publication by Socinfo 2019.}

\end{abstract}
\section{Introduction}
%Studying the stance in social media has gained a noticeable interest due to its valuable applications in event analysis, polarization detection and rumours detection \cite{borge2015content,garimella2018polarization,weber2013secular}. 
The stance can be defined as the expression of the individual's standpoint toward a proposition \cite{biber1988adverbial}. 
Detecting the stance towards an event is a sophisticated process where various factors play a role in discovering the viewpoint, including personal and social aspects. Most of the studies in this area have focused on using the textual elements of user’s posts such as sentiment of the text to infer the stance \cite{somasundaran2010recognizing,elfardy2016cu,ebrahimi_joint_2016}. While the goal of the stance detection is to determine the favorability towards a given entity or topic \cite{mohammad_stance_2017}, sentiment analysis aims to determine whether the emotional state of a given text is positive, negative, or neutral \cite{liu2010sentiment}. There is a rich body of research where the sentiment has been used solely to discover the viewpoints towards an event \cite{lee2018using,overbey2017linkin,unankard2014predicting,tsolmon2012extracting}. These studies expected that the sentiment polarity could indicate the stance. However, another line of research develops a stance specific model to infer the viewpoints where sentiment is being neglected \cite{kareem_2019,darwish_improved_2017,trabelsi2018unsupervised}. As the dependence on sentiment as a sole factor for the stance prediction has been found to be suboptimal, which might indicate a weak relation between sentiment and stance \cite{mohammad_stance_2017,elfardy2016cu}.  
%Using the ... add something about the other factors
%Factors such as user interactions, textual cues has been heavly investigated and yet there 

%found out that using sentiment as a proxy to predict the stance is sub-optimal \cite{mohammad_stance_2017}.  
Accordingly, it becomes important to examine the relation between the sentiment and the stance for viewpoint discovery toward an event. This leads us to pose the following research questions:
\begin{itemize}
%How does sentiment differ across the expressed stances?
%Can sentiment's polarity substitute the stance detection in inferring a viewpoint
  \item RQ1: Can sentiment polarity be used to capture the stance towards an event? % Answer the stacked columns showing that the overall sentiment failed to capture the real stance in the topics.
  %How proper is using sentiment instead of stance to denote a viewpoint?
  %What is the extent of the alignment between the sentiment and stance?
  \item RQ2: How does sentiment align with stance? When does positive/negative sentiment indicate support/against stance?% The heatmap shows the distribution of sentiment and stance. 
  
  %What is the extent of the correlation between sentiment and stance ?
\end{itemize}  
These questions aim to identify whether the sentiment can substitute the stance by studying the polarity nature of the expressed stance. In other words, this study examines whether the supporting/opposing stances can be identified with positive/negative sentiment. To answer these questions, we used the SemEval stance dataset \cite{mohammad_semeval-2016_2016}, the popular stance dataset that contains sentiment and stance labels. 
To further validate the results, we constructed a new stance detection dataset that has about 6000 tweets towards four topics and annotated with gold labels for sentiment and stance. This dataset contains the parent tweets along with reply tweets, which provides contextualized information for the annotator and helps in judging the sentiment and stance of the reply tweets. After that, we analyze the datasets to determine the degree of the correlation between sentiment polarity and the gold label stance. 
%We used the popular benchmark for stance detection, the SemEval stance dataset \cite{mohammad_semeval-2016_2016}, to validate the results.

%Another line of study designs stance detection model by using sentiment as feature to predict the stance toward a target of interest ~\cite{igarashi_tohoku_2016,krejzl_uwb_2016,ebrahimi_joint_2016,elfardy2016cu}. These studies hypothesised that sentiment analysis is closely related to stance detection. However, using sentiment as proxy to predict the stance tends to be sub-optimal \cite{mohammad_stance_2017}. Many studies have reached the conclusion that using sentiment was insignificant for the stance classifier \cite{elfardy2016cu,mohammad_stance_2017}.  
%The contribution of this paper is as follow:
%\begin{itemize}
 %   \item We construct a new contextualized dataset that has about 6000 tweets towards four topics and annotated with gold labels for sentiment and stance. %This dataset consist of the parent tweet along with reply tweet which provides further context to the annotated tweet.
  %\item We carryout an analysis to determine the degree of the correlation between sentiment polarity and stance. We used the popular Benchmark stance dataset to validate the results.
%\end{itemize}
 
%------------------------------------

\section{Related work}
In the literature, sentiment has been widely used either to infer the public opinion or as a factor to help in detecting the stance towards an event. The next sections illustrate these cases with a focus on studying the stance towards an event where the simple sentiment has been used either by using a sentiment lexicon or the textual polarity of the text.   
%1-sentiment as proxy to stance
%2- work defining the relation between sentiment and stance 
\subsection{Sentiment as stance}
Sentiment has been used interchangeably with stance to indicate the viewpoint detection \cite{park2011politics,hu2013listening,smith2017analyzing,lee2018using,unankard2014predicting,tsolmon2012extracting,agarwal2018geospatial}. In these studies the sentiment polarity has been used purely as the only factor to detect the viewpoint towards various events in social media. For instance, the work of \cite{smith2017analyzing} used sentiment to investigate the opinion towards the terrorist attack in Paris, during November 2015. They used annotators from Crowdflower to label the sentiment (negative, positive or neutral) as expressed in the tweet and used these labels as a way to analyse the public reaction toward Paris attack in 2015. %Their findings indicates that the proportion of negative English tweets exceeds the negative French tweets.  Finally they used the negative sentiment toward the event as way to gauge the public reaction as they compared the local ''French'' reaction to the non-local reaction ''English''.
In a study done by \cite{park2011politics}, they used the sentiment to discover the political leaning of the commenter on news articles. In their study, a sentiment profile constructed for each commenter to help in tracking their polarity toward a political party. For instance, a liberal commenter uses negative comments in conservative articles and  positive comments to liberal articles.

A more recent study by \cite{lee2018using} used the sentiment to examine the opinions following the release of James Comey’s letter to Congress before the 2016 US presidential election day. The previous study categorized 25 most common hashtags with sentiment polarity towards Hillary Clinton and Trump. Furthermore, the work of \cite{unankard2014predicting} used sentiment to analyze the political preferences of the users for the 2013 Australian federal election event. For the sentiment they recruited three annotators to label the tweet with a polarity score (positive, negative or neutral). In their study they used aspect-level sentiment for predicting user’s political preference and they overlooked the cases where the sentiment is negative and the stance is expressing a support viewpoint. 

Another study \cite{tsolmon2012extracting} developed an opinion score equation based on sentiment lexicon and frequency of a term to infer the users’ opinions towards events as they extracted from the timeline.
In addition, the work of \cite{hu2013listening} designed topic-sentiment matrix to infer the crowd’s opinion.  Another recent study by \cite{agarwal2018geospatial} used AFINN-111 dictionary for sentiment analysis and used sentiment polarity as an indication of the opinion towards Brexit. 
All of the above studies treated sentiment as the indicator of the stance toward the event of the analysis.

\subsection{Sentiment as proxy for stance} 
 Another line of research used sentiment as a feature to predict the stance \cite{somasundaran2010recognizing,elfardy2016cu,ebrahimi_joint_2016,mohammad_stance_2017}. In the popular SemEval stance dataset \cite{mohammad_semeval-2016_2016}, the tweets are labeled with sentiment and stance to provide a public benchmark to evaluate the stance detection systems. In their work, they showed that using sentiment features are useful for stance classification when they combined with other features  and not used alone. 
 The work of \cite{ebrahimi_joint_2016} used an undirected graphical model that leverages interactions between sentiment and the target of stance to predict the stance. Also, the work of \cite{somasundaran2010recognizing} developed a stance classifier that used sentiment and arguing expressions by using sentiment lexicon along with arguing lexicon which outperforms Uni-gram features system. In \cite{igarashi_tohoku_2016} they used SentiWordNet to produce sentiment for each word and use the sentiment value along with other features to predict the stance in SemEval stance dataset and compared with CNN stance model. They found that feature based model performed better in detecting stance. The work of \cite{krejzl_uwb_2016} used surface-level, sentiment and domain-specific features to predict the stance on SemEval stance dataset. Overall, the use of sentiment in conjunction with other features helps in predicting the stance but not as the only dependent feature. 
 %The work of \cite{sobhani2016detecting} showed that using sentiment features are useful for stance classification when they combined with other features  and not used alone.
 %The below study is the strange one from the unknown journal 
 %A more recent study by \cite{sun2019stance} developed a joint model between sentiment and stance using shared LSTM layer to learn the shared  representation  between stance  and  sentiment. Their model leverages the auxiliary representation to assist the performance of the  stance detection task. 
 
 The work of \cite{mohammad_stance_2017,sobhani2016detecting} studied the extent to which the sentiment is correlated with the stance in the sense of enhancing stance classifier. The main focus of the previous study was to investigate the best features for the stance classification model. In their work, they concluded that sentiment might be beneficial for stance classification, but when it is combined with other factors. 
 
 This study investigates another dimensionality of the sentiment-stance relation with focus on gauging the alignment between sentiment and stance by analysing in depth the relation of how the stance is being expressed in conjunction with the sentiment. 

% https://www.aclweb.org/anthology/W14-2109
\begin{table}[!t]
\centering
\small
\begin{tabular}{l|c||l|c} 
  \hline
  \textbf{SemEval stance}  & \# &\textbf{CD stance}  & \#  \\
  \hline
  Atheism (A) &  733 & Antisemitic (AS) &  1050\\
   Climate Change is Concern (CC) & 564 &Gender (G) & 1050 \\
   Feminist Movement (FM) & 949 & Immigration (I)&  3174  \\
   Hillary Clinton (HC) & 934 & LGBTQ (L) &  1050\\
   Legalization of Abortion (LA) & 883 & &  \\
  \hline
  Total & 4063 & Total &6324 \\
    %-------------------------------------------
  \hline
\end{tabular}
\caption{Number of tweets for each topic. }
\label{tab:datasets}
\squeezeup
\end{table}
\squeezeup

\section{Data collection}
We study the sentiment nature in the expressed stance. To accomplish this, we used SemEval stance dataset which contains about 4000 tweets on five topics, including Atheism (A), Climate Change (CC), the Feminist Movement (FM), Hillary Clinton (HC) and the Legalisation of Abortion (LA). Furthermore, we designed a context-dependent (CD) stance dataset that contains  6324 reply tweets covering four controversial topics: Antisemitic (AS), Gender (G), Immigration (I), LGBTQ (L). Table \ref{tab:datasets} shows the distribution of the tweets with respect to each topic. In this dataset, each tweet has been annotated by five annotators using Figure-eight platform \footnote{https://figure-eight.com/}, and the label with a majority vote is assigned. We used the same annotations guideline of SemEval stance dataset \cite{mohammad_semeval-2016_2016}. Since CD dataset is all reply tweets, the parent tweet along with reply tweet has been provided to the annotators to understand the context of the conversation to better judge the sentiment and stance. 

\section{Methodology}
\subsection{Analysis of the correlation patterns}

\begin{figure*}[t]
    \centering
    %Font size 
   % \setkeys{Gin}{width=0.40\textwidth}
    \setkeys{Gin}{width=0.49\textwidth}
    %imgs/stance_colum.pdf
\subfloat[ SemEval stance
          \label{fig:subfig-a}]{\includegraphics{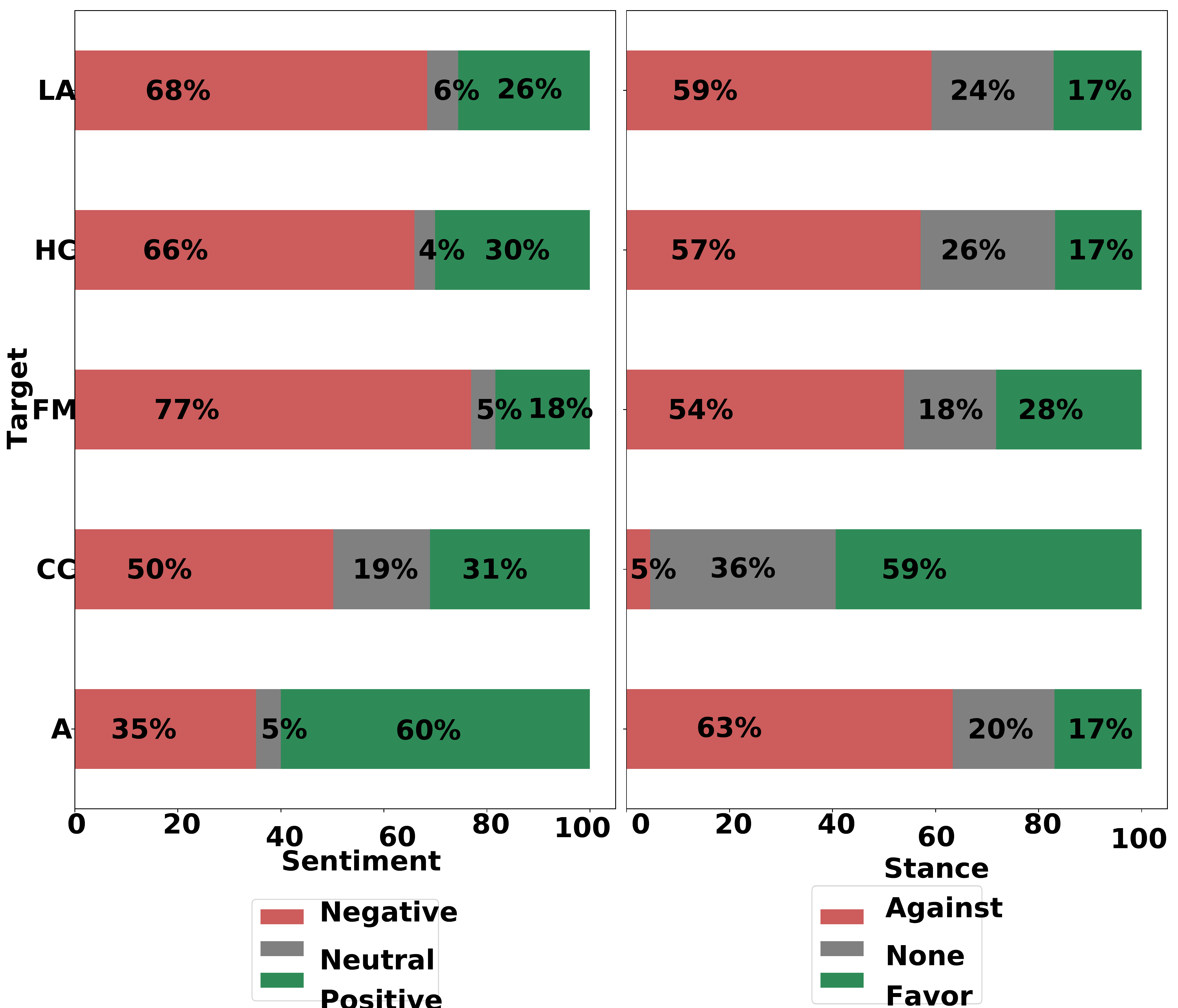}}
    %\hfill
    \hspace{0.5pt}
    %imgs/stance_sentiment_colum_new.pdf
\subfloat[ CD stance
          \label{fig:subfig-b}]{\includegraphics{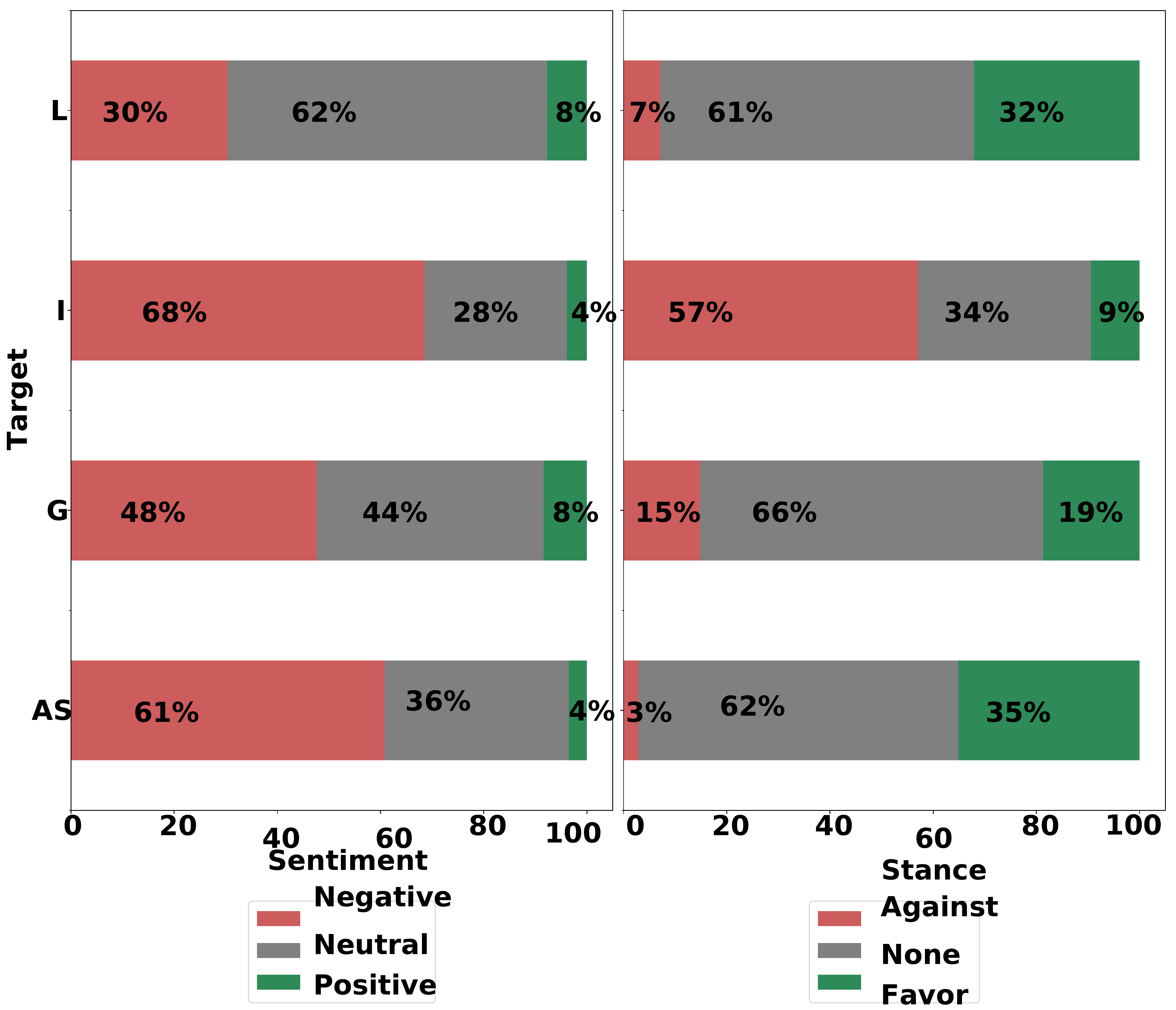}}
\caption{The distribution of sentiment and stance with respect to each topic. }
\label{fig:cloumn_bar_dist}
    \end{figure*}

To get a good insight of how the stance is being expressed, we first analyze the distribution of stance and sentiment on the topic level. Figures \ref{fig:cloumn_bar_dist} a and b, illustrate the stance and sentiment distribution in the SemEval stance dataset and CD stance dataset, respectively.  Overall the negative sentiment constitutes the major polarity of the most topics. This reveals the tendency of using negative sentiments to express a viewpoint in a controversial topic.  It can be observed that for the climate change the supporting stance constitutes about 59\%; however the overall tweets with negative sentiment constitute 50\%. Furthermore, 30\% of the LGBTQ tweets show negative sentiment, while only 7\% of the tweets express the opposing stance. From these numbers, it is clear that sentiment does not simply represent stance.

%----------------------------
%----------------------------

\begin{figure*}[t]
    \centering
   % \setkeys{Gin}{width=0.40\textwidth}
    \setkeys{Gin}{width=0.38\textwidth}
\subfloat[ SemEval stance
          \label{fig:subfig-a}]{\includegraphics{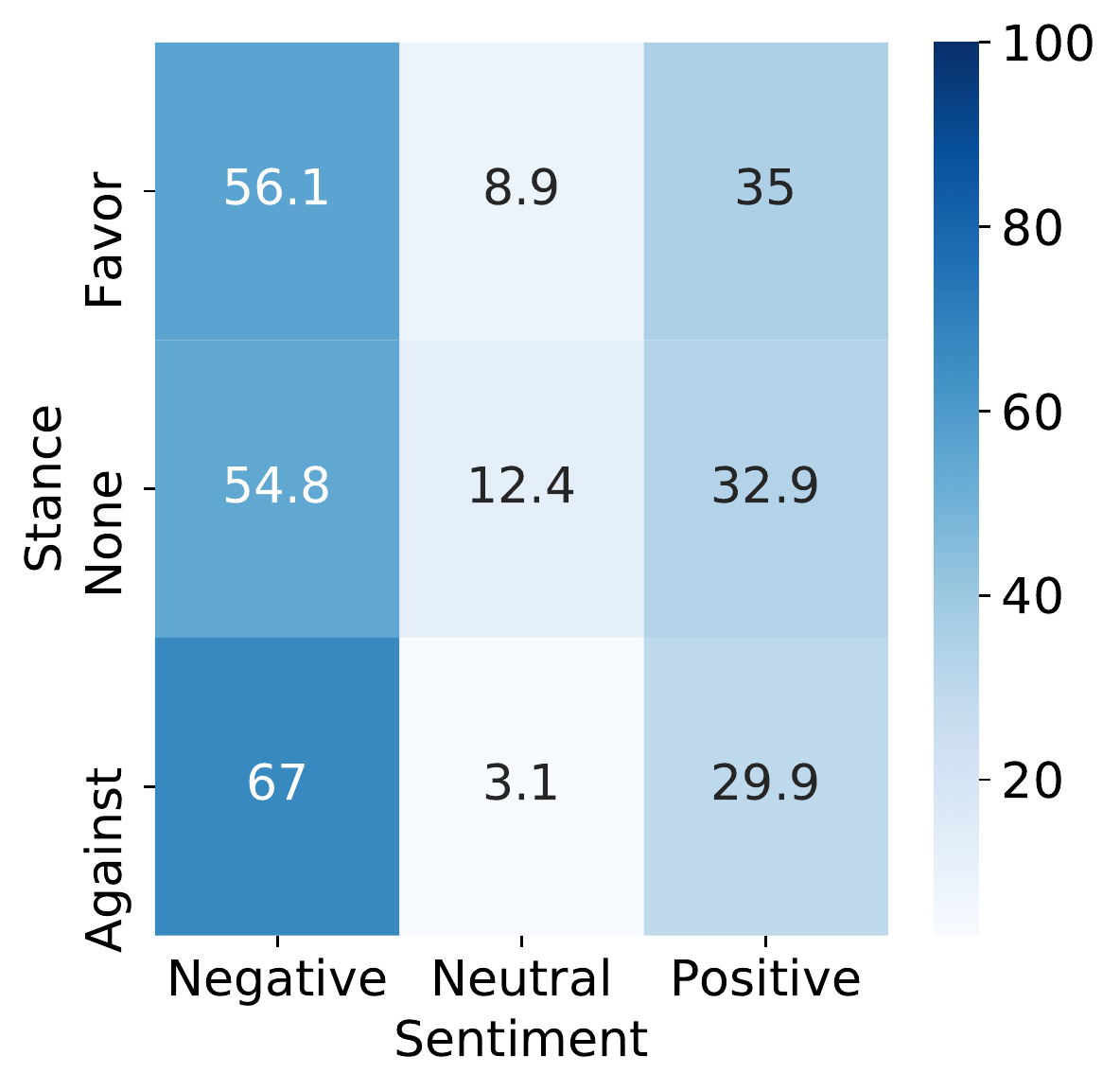}}
    %\hfill
\subfloat[ CD stance
          \label{fig:subfig-b}]{\includegraphics{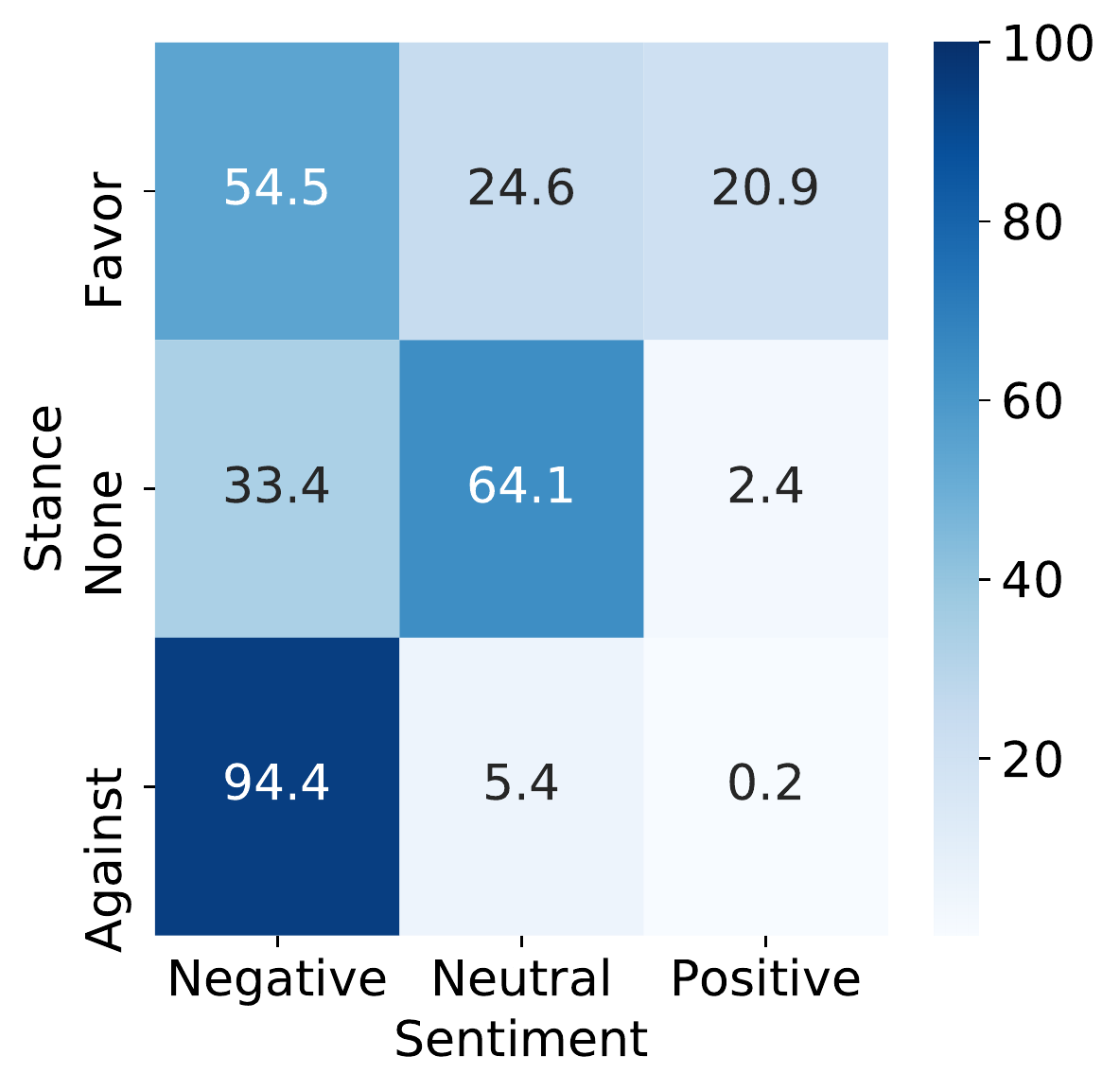}}
\caption{Distribution of sentiment per a given stance.  }
\label{fig:heatmap}
\vspace{-4mm}
    \end{figure*}

%-------------------------Heatmap
Figure~\ref{fig:heatmap} illustrates the sentiment distribution over the stance in the two datasets. The graphs show that the negative sentiment constitutes the major polarity over the Favor and Against stances. As the negative sentiment represents over 56\% and 54\% of the supporting stance in the SemEval and CD stance datasets, respectively. These results reveal the tendency of using negative sentiments to express a viewpoint towards a controversial topic.  

 %----heatmap for the distribution of 

%-------table unique users with same stance
\begin{comment}
\begin{table}
\centering
\begin{tabular}{l|ccc} 
  \hline
  \textbf{dataset}  & Against-negative & positive-favor & overall  \\
  \hline
  SemEval stance& 699& 273& 2263 \\
  CD stance& 1407&206 &4459\\
  %-------------------------------------------
  \hline
\end{tabular}
\caption{Number of unique users with similar stance-sentiment. }
\label{tab:resultsall}
\end{table}
\end{comment}
%RQ#1 : Can sentiment polarity be used to capture the stance towards an event

Table \ref{table:Sentiments_Example} shows some examples where the sentiment does not reflect the stance. Examples 1 and 2 show tweets with an opposing viewpoint to targets, while using positive sentiment. Examples 3 and 4 show the opposite situation, where the expressed stance is supporting, while the sentiment is negative. 

These results show that sentiment fails to detect the real stance toward a topic. There is a clear mismatching between the negative/positive sentiment and the supporting/against stance. Even with the dominance of the negative sentiment in most of the topics, yet the overall stance has shown a mixer of support viewpoint. 
%This gives an insight of the behavioural pattern in the expressing the viewpoint and the tone used. 

%%--------------------------------------
\begin{table}[t]
\centering
\small
\begin{tabular} { >{\centering\arraybackslash}m{0.01\linewidth}  >{\arraybackslash}m{0.70\linewidth} >{\centering\arraybackslash}m{0.09\linewidth} >{\centering\arraybackslash}m{0.09\linewidth} >{\centering\arraybackslash}m{0.09\linewidth} }
  \hline
   \# & Tweet & Target & Sent. & Stance  \\
  \hline
  1 & Life is our first and most basic human right. & LA& + & Against \\
  2 & @realDonaldTrump Thank you for protecting our border & I & + & Against\\
  3 & The biggest terror threat in the World is climate change \#drought \#floods & CC & - &  Favor\\
  4 & In the big picture, religion is bad for society because it blunts reason.  \#freethinker & A & - & Favor\\
  %4 & You are NOT a feminist if you are a racist. If you don't stand for all girls of every color, shape, size then u ain't a feminist & Feminist Movement & - & Favor\\
  \hline
  %-------------------------------------------
 \hline
\end{tabular}
\caption{Differences between sentiment and stance. Targets: Legalization of Abortion (LA), Immigration (I), Atheism (A), Climate Change (CC).}
\label{table:Sentiments_Example}
\squeezeup
\end{table}
%-----------------------------  

\squeezeup

\section{Discussion}
% RQ1: RQ1: Can sentiment's polarity substitute the stance detection in inferring a viewpoint ?
Our first research question concerns with whether the sentiment captures the real stance, can be answered with dissenting. The previous analysis shows that the sentiment cannot substitute the stance in general. The words choice gap exists for in-favor stance and positive sentiment (Appendix A). Subsequently, We noticed that sentiment has failed to discover the public opinion towards most of the topics in the two datasets. Hence, using the sentiment polarity as the only factor to predict the public opinion potentially leads to misleading results. The result of the mismatch between in-favor and positive stance was sizable. The positive sentiment failed to distinguish the supporter viewpoints. 
%RQ2:  What is the extent of the alignment between the sentiment and stance?

 As for the overall alignment between the sentiment and stance, there is a noticeable disparity between sentiment and stance for a given topic. In general, the sentiment tends to be negative in the expressed stance as a way to rebuttal or defend the viewpoint and show support or opposing stance.  The negative sentiment could help in discovering some of the against stances, but it will be mixed with a proportion of the supporter viewpoints. 
 
 In summary, our analysis in this paper illustrates the sophisticated nature of stance detection and that it cannot be simply captured using the sentiment polarity. This finding is crucial, especially when assessing the credibility of results in studies that used sentiment to measure public support of a given topic on social media.
%We arial events using only sentiment an indication of support.gue or soc that sentiment as proxy help in improving the identification of the against stance. As in semeval dataset the overall against stance constitutes about 54\% for four topics and 5\% for climate change this is the reason that most of studies that used sentiment solely failed to predict the stance in this topic. 
%Furthermore, negative sentiment tends to be the prevailing theme in the expressed stance as a way to rebuttal or defend the viewpoint and show support or opposing stance.

\section{Conclusion}
In this paper, we study the relation between the sentiment and the stance. To gauge the extent of this relation, we constructed a new stance dataset with gold sentiment and stance labels. Then we conducted a textual and quantitative analysis of the expressed stance with respect to the sentiment polarity. Our study provides evidence that sentiment cannot substitute the stance. As a final consideration, researcher should be more cautious when it comes to identifying the viewpoints toward an event and to take into account the clear difference between the sentiment and the stance. As using sentiment purely overshadows the real stance and leads to truncated results. 

%Future direction of this research is to conduct additional analysis to the semantic differences to further gauge the similarity between the words choice. 
% I am not planning to do it but since this is a short paper then their should be a future direction.
 
%\begin{theorem}
%This is a sample theorem. The run-in heading is set in bold, while
%the following text appears in italics. Definitions, lemmas,
%propositions, and corollaries are styled the same way.
%\end{theorem}
%
% the environments 'definition', 'lemma', 'proposition', 'corollary',
% 'remark', and 'example' are defined in the LLNCS documentclass as well.
%
%\begin{proof}
%Proofs, examples, and remarks have the initial word in italics,
%while the following text appears in normal font.
%\end{proof}
%
% ---- Bibliography ----
%
% BibTeX users should specify bibliography style 'splncs04'.
% References will then be sorted and formatted in the correct style.
%
 \bibliographystyle{splncs04}
 \bibliography{ms}

\begin{thebibliography}{10}
\providecommand{\url}[1]{\texttt{#1}}
\providecommand{\urlprefix}{URL }
\providecommand{\doi}[1]{https://doi.org/#1}

\bibitem{achananuparp2008evaluation}
Achananuparp, P., Hu, X., Shen, X.: The evaluation of sentence similarity
  measures. In: International Conference on data warehousing and knowledge
  discovery. pp. 305--316. Springer (2008)

\bibitem{agarwal2018geospatial}
Agarwal, A., Singh, R., Toshniwal, D.: Geospatial sentiment analysis using
  twitter data for uk-eu referendum. Journal of Information and Optimization
  Sciences  \textbf{39}(1),  303--317 (2018)

\bibitem{an2019political}
An, J., Kwak, H., Posegga, O., Jungherr, A.: Political discussions in
  homogeneous and cross-cutting communication spaces (2019)

\bibitem{biber1988adverbial}
Biber, D., Finegan, E.: Adverbial stance types in english. Discourse processes
  \textbf{11}(1),  1--34 (1988)

\bibitem{darwish_improved_2017}
Darwish, K., Magdy, W., Zanouda, T.: Improved {Stance} {Prediction} in a {User}
  {Similarity} {Feature} {Space}. In: {ASONAM}'17 (2017)

\bibitem{ebrahimi_joint_2016}
Ebrahimi, J., Dou, D., Lowd, D.: A {Joint} {Sentiment}-{Target}-{Stance}
  {Model} for {Stance} {Classification} in {Tweets}. In: {COLING}. pp.
  2656--2665 (2016)

\bibitem{elfardy2016cu}
Elfardy, H., Diab, M.: Cu-gwu perspective at semeval-2016 task 6: Ideological
  stance detection in informal text. In: Proceedings of the 10th International
  Workshop on Semantic Evaluation (SemEval-2016). pp. 434--439 (2016)

\bibitem{gomaa2013survey}
Gomaa, W.H., Fahmy, A.A.: A survey of text similarity approaches. International
  Journal of Computer Applications  \textbf{68}(13),  13--18 (2013)

\bibitem{hu2013listening}
Hu, Y., Wang, F., Kambhampati, S.: Listening to the crowd: automated analysis
  of events via aggregated twitter sentiment. In: Twenty-Third International
  Joint Conference on Artificial Intelligence (2013)

\bibitem{igarashi_tohoku_2016}
Igarashi, Y., Komatsu, H., Kobayashi, S., Okazaki, N., Inui, K.: Tohoku at
  {SemEval}-2016 {Task} 6: {Feature}-based {Model} versus {Convolutional}
  {Neural} {Network} for {Stance} {Detection}. In: {SemEval}@ {NAACL}-{HLT}.
  pp. 401--407 (2016)

\bibitem{kareem_2019}
Kareem, D., Peter, S., Aupetit, M.J., Preslav, N.: Unsupervised user stance
  detection on twitter. arXiv preprint arXiv:1904.02000  (2019)

\bibitem{krejzl_uwb_2016}
Krejzl, P., Steinberger, J.: {UWB} at {SemEval}-2016 {Task} 6: {Stance}
  {Detection}. In: {SemEval}@ {NAACL}-{HLT}. pp. 408--412 (2016)

\bibitem{lee2018using}
Lee, H.W.: Using twitter hashtags to gauge real-time changes in public opinion:
  An examination of the 2016 us presidential election. In: International
  Conference on Social Informatics. pp. 168--175. Springer (2018)

\bibitem{liu2010sentiment}
Liu, B., et~al.: Sentiment analysis and subjectivity. Handbook of natural
  language processing  \textbf{2}(2010),  627--666 (2010)

\bibitem{mohammad_semeval-2016_2016}
Mohammad, S., Kiritchenko, S., Sobhani, P., Zhu, X.D., Cherry, C.:
  {SemEval}-2016 {Task} 6: {Detecting} {Stance} in {Tweets}. In: {SemEval}@
  {NAACL}-{HLT}. pp. 31--41 (2016)

\bibitem{mohammad_stance_2017}
Mohammad, S.M., Sobhani, P., Kiritchenko, S.: Stance and sentiment in tweets.
  ACM Transactions on Internet Technology (TOIT)  \textbf{17}(3), ~26 (2017)

\bibitem{overbey2017linkin}
Overbey, L.A., Batson, S.C., Lyle, J., Williams, C., Regal, R., Williams, L.:
  Linking twitter sentiment and event data to monitor public opinion of
  geopolitical developments and trends. In: International Conference on Social
  Computing, Behavioral-Cultural Modeling and Prediction and Behavior
  Representation in Modeling and Simulation. pp. 223--229. Springer (2017)

\bibitem{park2011politics}
Park, S., Ko, M., Kim, J., Liu, Y., Song, J.: The politics of comments:
  predicting political orientation of news stories with commenters' sentiment
  patterns. In: Proceedings of the ACM 2011 conference on Computer supported
  cooperative work. pp. 113--122. ACM (2011)

\bibitem{smith2017analyzing}
Smith, K.S., McCreadie, R., Macdonald, C., Ounis, I.: Analyzing
  disproportionate reaction via comparative multilingual targeted sentiment in
  twitter. In: Proceedings of the 2017 IEEE/ACM International Conference on
  Advances in Social Networks Analysis and Mining 2017. pp. 317--320. ACM
  (2017)

\bibitem{sobhani2016detecting}
Sobhani, P., Mohammad, S., Kiritchenko, S.: Detecting stance in tweets and
  analyzing its interaction with sentiment. In: Proceedings of the Fifth Joint
  Conference on Lexical and Computational Semantics. pp. 159--169 (2016)

\bibitem{somasundaran2010recognizing}
Somasundaran, S., Wiebe, J.: Recognizing stances in ideological on-line
  debates. In: Proceedings of the NAACL HLT 2010 Workshop on Computational
  Approaches to Analysis and Generation of Emotion in Text. pp. 116--124.
  Association for Computational Linguistics (2010)

\bibitem{trabelsi2018unsupervised}
Trabelsi, A., Zaiane, O.R.: Unsupervised model for topic viewpoint discovery in
  online debates leveraging author interactions. In: Twelfth International AAAI
  Conference on Web and Social Media (2018)

\bibitem{tsolmon2012extracting}
Tsolmon, B., Kwon, A.R., Lee, K.S.: Extracting social events based on timeline
  and sentiment analysis in twitter corpus. In: International Conference on
  Application of Natural Language to Information Systems. pp. 265--270.
  Springer (2012)

\bibitem{unankard2014predicting}
Unankard, S., Li, X., Sharaf, M., Zhong, J., Li, X.: Predicting elections from
  social networks based on sub-event detection and sentiment analysis. In:
  International Conference on Web Information Systems Engineering. pp. 1--16.
  Springer (2014)

\end{thebibliography}
\begin{appendix}
\section{Analysis of the textual patterns}
To gauge the similarity between the vocabulary choice that has been used to express the sentiment and stance we analyzed the tweets in the two datasets using Jaccard similarity. We used Jaccard coefficient the widely adopted measure to capture the overlap between two sets \cite{an2019political,achananuparp2008evaluation,gomaa2013survey}.  In this analysis, for each sentiment and stance gold labels we combine all tweets and use Term Frequency-Inverse Document (TF-IDF), to find important words in each type of sentiment and stance. In order to compute the TF-IDF on tweet level we consider each tweet as document. Using TF-IDF helps in filtering out less significant words. 
The Jaccard similarity between the set of sentiment and stance words defined as following:

\begin{equation}
\small
Jaccard(W_{sentiment},W_{stance}) =   \frac{W_{sentiment} \cap W_{stance}  }  {W_{sentiment} \cup W_{stance}}
\end{equation}
%Comparing the similarity of the words that used to convey a stance and sentiment in the two data-sets Semeval stance and CD stance.

Where W$_{sentiment}$ and W$_{stance}$ denote the list of top N words by TF-IDF value for the tweets with specific sentiment and stance type. 

%------------------------------------------
%’’Correlation does not equal causation‘’!
%------------------Donuts chart 
\begin{figure*}
    \centering
   % \setkeys{Gin}{width=0.40\textwidth}
   %old common word file: neg_against_pos_fav_semeval.pdf
   %tfidf_semeval.pdf
    \setkeys{Gin}{width=0.4\textwidth}
\subfloat[ Semeval stance
          \label{fig:subfig-a}]{\includegraphics{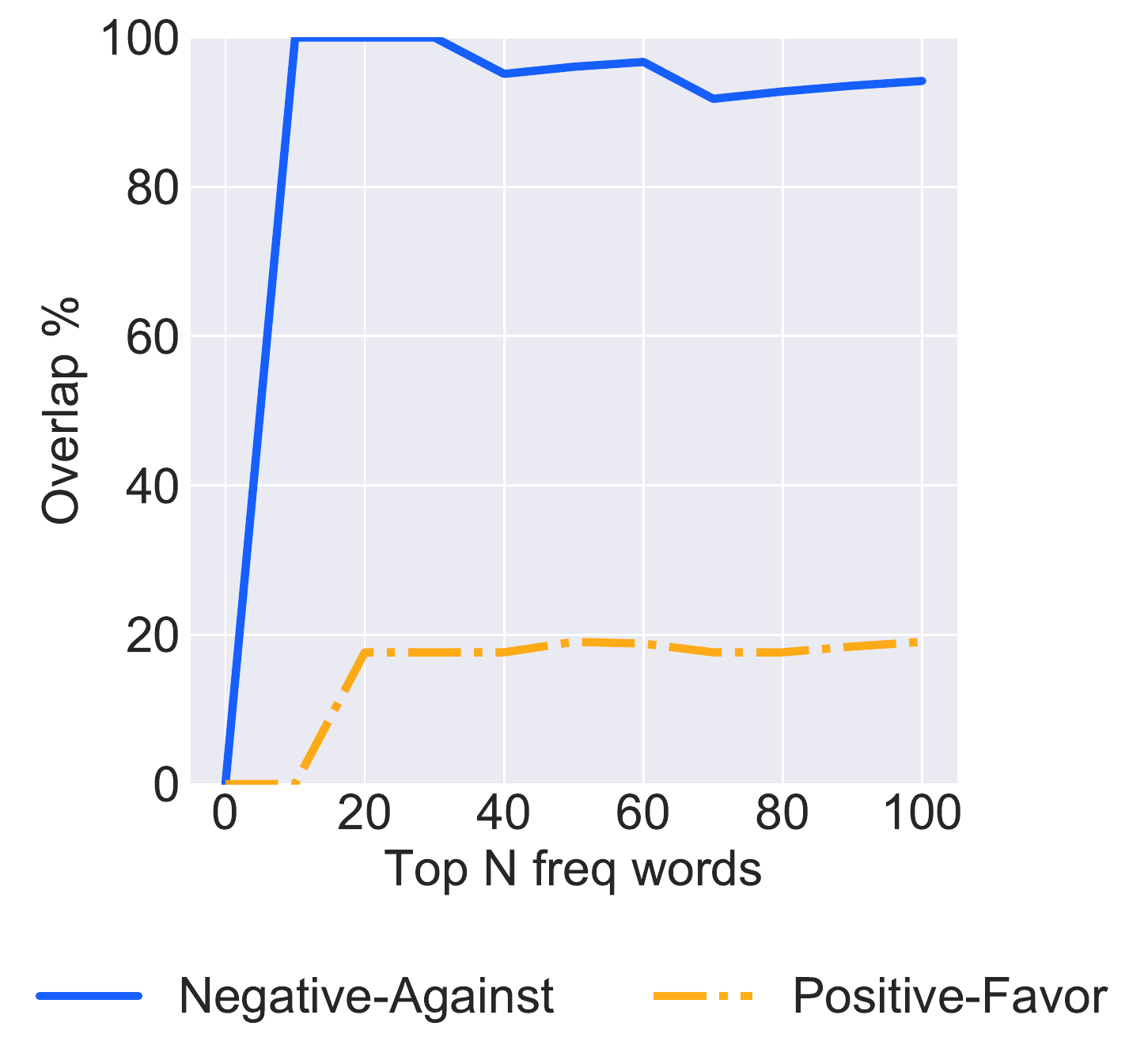}}
    %\hfill
    %tfidf_cd.pdf
    %file for common words:neg_against_pos_fav_newdataset.pdf
\subfloat[ CD stance
          \label{fig:subfig-b}]{\includegraphics{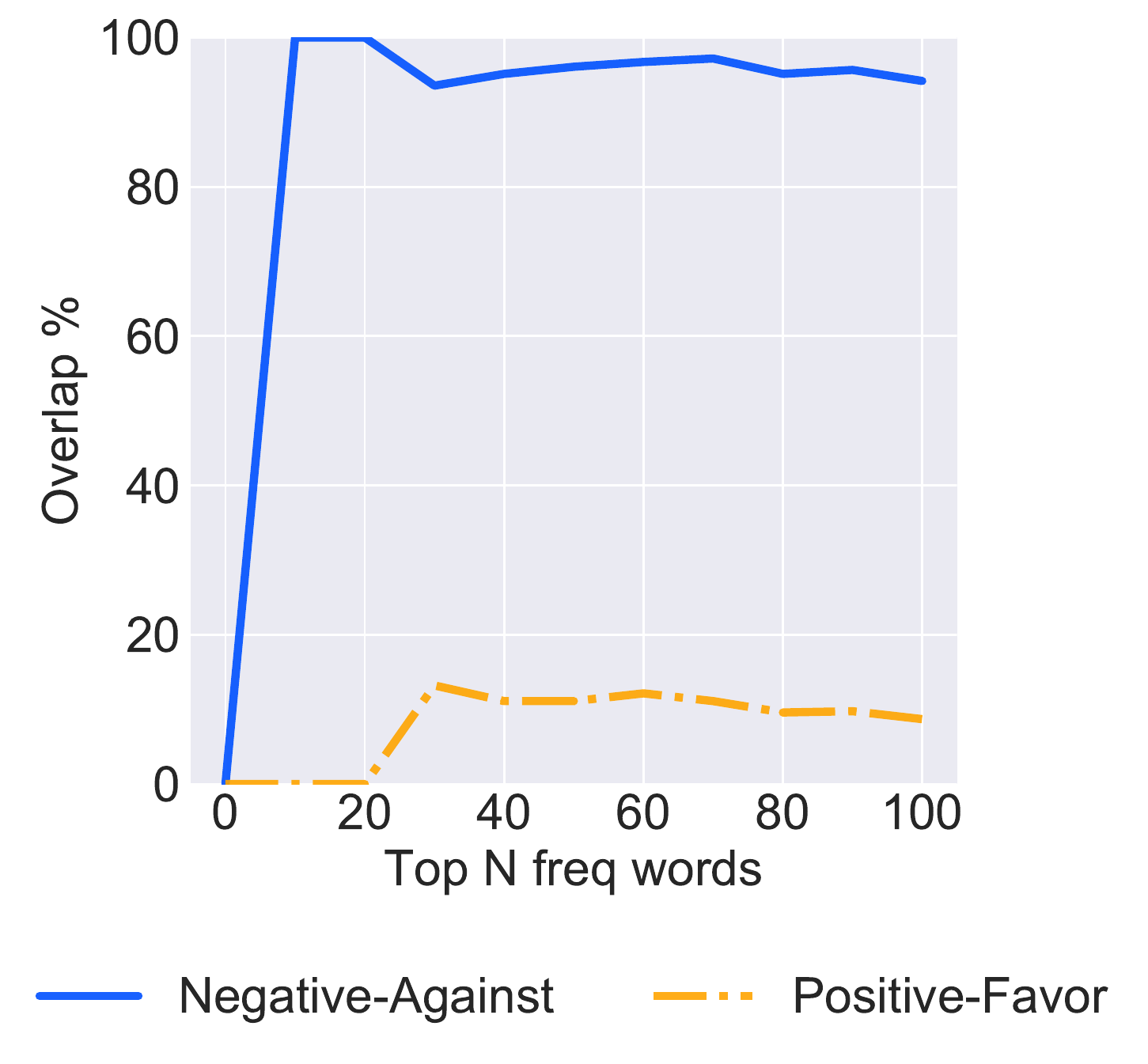}}
\caption{Jaccard similarity of the top N-most frequent words between sentiment and stance. }
\label{fig:Jaccard_similarity}
\squeezeup
    \end{figure*}
\begin{figure*}[t]
 \squeezeup
    \centering
      \setkeys{Gin}{width=0.5\textwidth}

    %\hfill
\subfloat[ SemEval stance
          \label{fig:subfig-b}]{\includegraphics{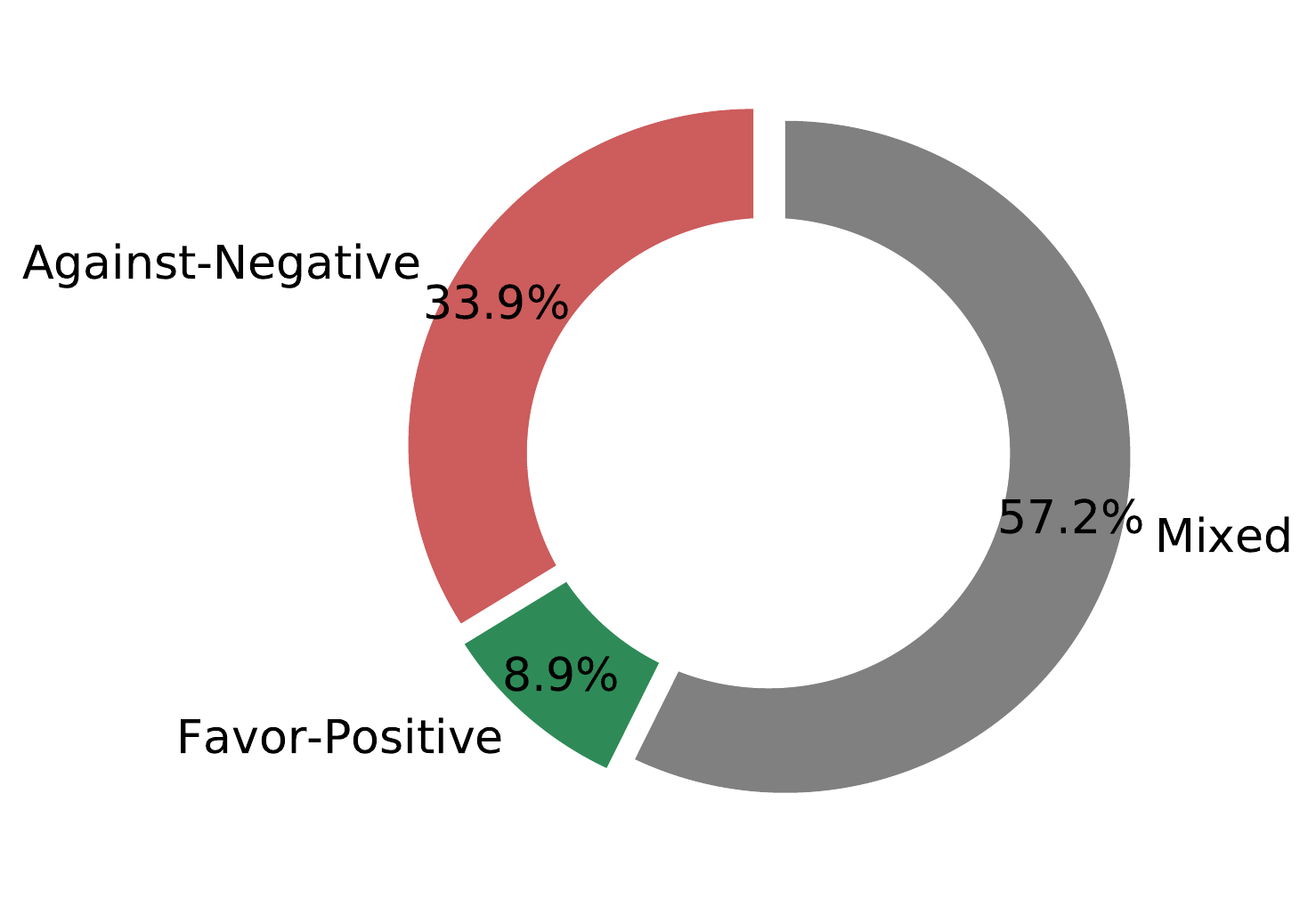}}
          %-----------------------
\subfloat[ CD stance
          \label{fig:subfig-a}]{\includegraphics{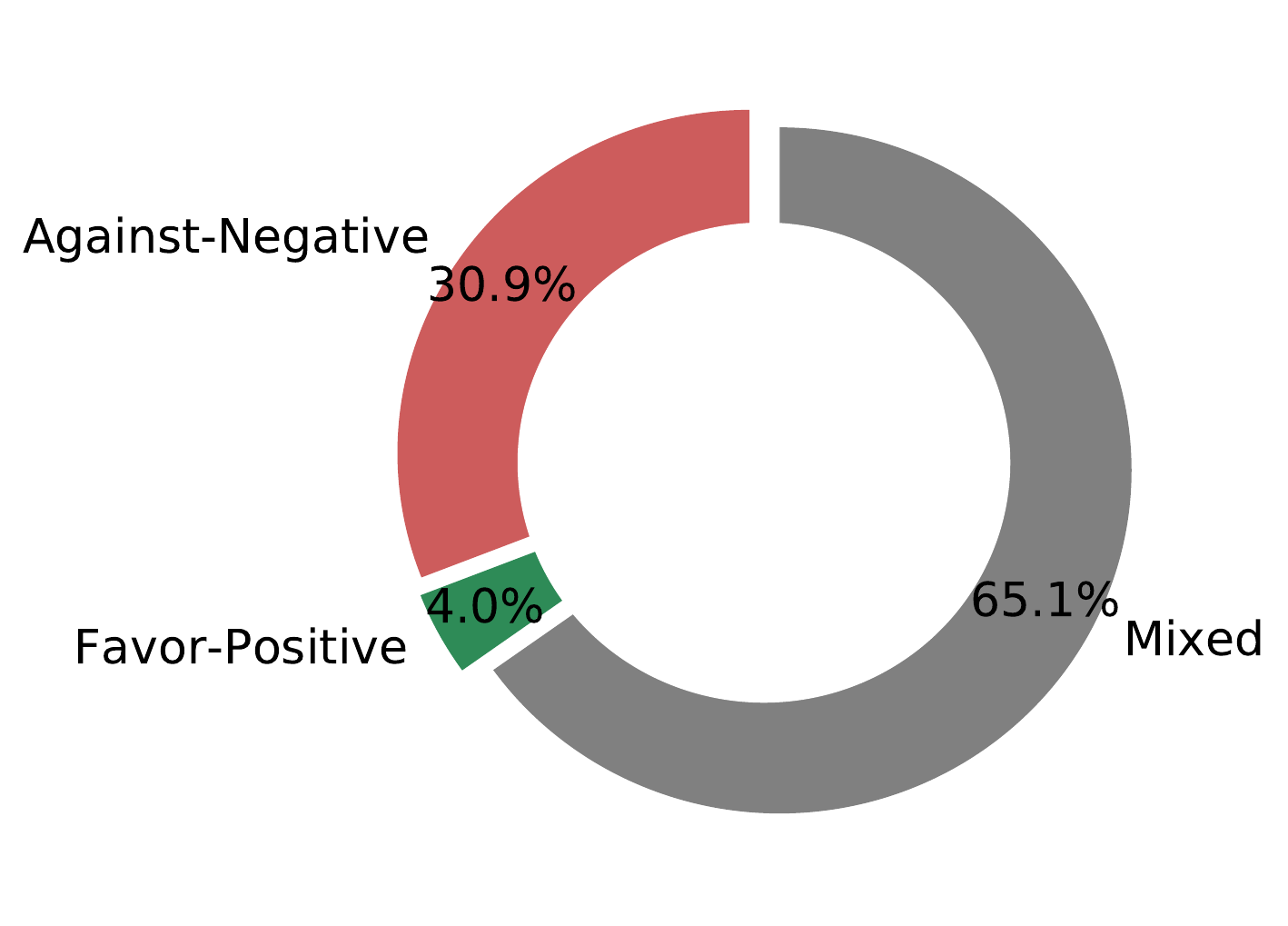}}          
\caption{Tweets with matching and mixed stance and sentiment. }
\label{fig:Donut:)}
\vspace{-4mm}
\squeezeup
    \end{figure*}

Fig \ref{fig:Jaccard_similarity} shows that the similarity between the words that have been used to express favor stance has less than 20\% of similarity with tweets that has a positive sentiment. That means users tend to express their Favor stance without using positive sentiment words. In contrast, the common words for against stance have the most significant similarity with against sentiment words. The Jacquard similarity become stable with growing N. As Fig \ref{fig:Donut:)} shows that the overall agreement between the sentiment and the stance is minuscule in general. The tweets that have against-negative labels constitutes less than 33\%. Similarly less than 8\% of the data has positive sentiment and favor stance. 
%In general, the tweets that have similar stance and sentiment constitute about 33.9\% and 30.9\% of the SemEval stance and CD dataset, prospectively.
This shows that in general negative words tend to be similar to the against words while the matching cases are minuscule. On the other-hand, the matching cases where the tweet express favor and positive sentiment constitute about 8.9\% and 4\% of the overall data of SemEval stance and CD stance dataset. 

\squeezeup

\end{appendix}

\end{document}